# Nature of nematic to twist bend nematic phase transition

Damian Pociecha,[a] Catriona Crawford,[b] Daniel A. Paterson,[b] John M. D. Storey,[b] Corrie T. Imrie,[b] Nataša Vaupotič,[c,d] Ewa Gorecka[a]

[a] University of Warsaw, Department of Chemistry, ul. Zwirki i Wigury 101, 02-089 Warsaw, Poland.
[b] Department of Chemistry, King's College, University of Aberdeen, Aberdeen, AB24 3UE, UK.
[c] Department of Physics, Faculty of Natural Sciences and Mathematics, University of Maribor, Koroška 160, 2000 Maribor, Slovenia
[d] Jožef Stefan Institute, Jamova 39, 1000 Ljubljana, Slovenia

**Abstract:** The critical behavior at the transition from uniform nematic to twist-bend modulated nematic phase is revealed and shown to be well explained by the mean field approximation. The study was performed on a group of materials that exhibit an unusually broad temperature range of the nematic phase above the twist-bend modulated nematic phase, so the critical range in which the order parameter fluctuations are strong and can be experimentally observed is wide. The formation of instantaneous helices is observed already several degrees above the transition temperature, strongly influencing the birefringence of the system. The analysis of a critical part of the birefringence changes provides a set of critical exponents that are consistent with the mean field approximation, indicating a large correlation length of helical fluctuations in the nematic phase.

Although rod-like molecules remain the most studied class of liquid crystalline materials, recent decades have seen an increasing interest in bent-core mesogens, stimulated by the discovery of their ability to form ferroelectric and antiferroelectric smectic phases [1]. There were also predictions that the nematic phase comprised of bent-core molecules may have some unusual properties. Initially, experimental studies focused on the search for the biaxial nematic phase [2,3] in which not only the long molecular axes but also their short axes show some degree of orientational order. The reports of the discovery of biaxial nematic phases consisting of bent core molecules turned out to be premature, [4,5]. However, this intense research resulted in the discovery of a new type of nematic phase: the twist-bend nematic ($N_{TB}$) – a phase with a short-pitch heliconical structure [6-9]. The $N_{TB}$ phase is a rare example of a chiral structure made from achiral molecular building blocks and was the first example of chiral symmetry breaking in a system with no long-range positional order. Despite an intensive research, many properties of the $N_{TB}$ phase still have to be understood. Dozov suggested that the spontaneous helical deformation of the director in the $N_{TB}$ phase is driven by the bend elastic constant becoming negative [10], either due to pure steric reasons or as a result of flexoelectric renormalization [11,12]. Experimentally, in the nematic phase of some bent-core materials a very small value of the bend elastic constant ($K_{33}$) was indeed measured [13], much smaller than the splay elastic constant ($K_{11}$) with the ratio of $K_{11}/K_{33}$ as high as 21 [14]. In addition, $K_{33}$ decreased with decreasing temperature. Such a behavior is in contrast to that found in the rod-like mesogens, for which usually $K_{33} > K_{11}$ and $K_{33}$ increases on cooling, following the square of the orientation order parameter ($S^2$) [15,16]. Interestingly, precise measurements made using a dynamic light scattering (DLS) [14] showed that in a narrow temperature range of the nematic phase close to the $N_{TB}$ phase, $K_{33}$ (and also the twist elastic constant $K_{22}$) drastically increases due to fluctuations, as locally formed helices constrain the macroscopic twist and bend deformations. However, there is still not sufficient data available to determine whether such a behavior is universal at the N – $N_{TB}$ phase transition. Specifically, the instantaneous formation of a helical structure in the nematic phase should also be detectable by precise birefringence measurements, allowing for a more detailed information about the phase transition between the uniform and heliconical phases to be obtained.

The heliconical nematic phase is described by a space variation of a two component vector and therefore the transition from a uniform nematic to a modulated nematic phase could be of the second order and should belong to the 3D XY universality class. However, verification of this hypothesis is difficult as the critical range in which the order parameter fluctuations are strong and can be experimentally observed is small, and for all the so far studied materials, the N – $N_{TB}$ phase transition remains of the first order [17]. Fluctuations should become more pronounced for materials having a wider nematic temperature range above $N_{TB}$, because the strength of the first order transition is related to the value of the effective twist elastic constant that becomes less negative if the temperature range of the nematic phase preceding the $N_{TB}$ phase increases [18]. To date, the only information about the critical behavior at the N – $N_{TB}$ phase transition is based on the heat capacity measurements for few compounds, that yielded the critical exponent $\alpha \sim 0.5$ in the $N_{TB}$ phase and no heat capacity anomalies above transition temperature [9, 17, 18], as predicted by the mean field model [19].

In this letter, we present the elastic and optical properties of asymmetric CB6O.Om dimers that show the $N_{TB}$ phase below a broad range (of approximately 40 K) of the nematic phase [20]. These materials exhibit an unusually strong optical birefringence anomaly above the N – $N_{TB}$ phase transition, which enables verification of the nature of the N – $N_{TB}$ phase transition. The results are compared with those obtained for the dimers CB6OCB and CB7CB.

The molecular structure and phase transition temperatures for the studied CB6O.Om compounds with different alkyl terminal chain lengths ($m = 4,5,8$) and the reference material CB6OCB are given in Fig. 1. The properties of the materials were studied by optical and dielectric methods, measurements were carried out for samples prepared in 1.6 – 3 µm thick glass cells, having ITO electrodes and a surfactant layer assuring planar alignment.

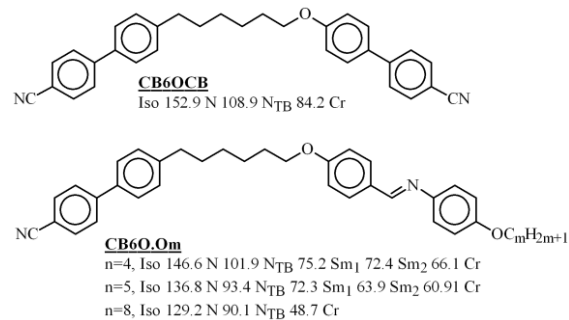

*Figure 1*. Molecular structures and phase transition temperatures (detected on cooling) for the studied compounds.

Birefringence was measured with a setup based on a photoelestic modulator (PEM-90, Hinds) working at a modulation frequency $f = 50$ kHz; as a light source a halogen lamp (Hamamatsu LC8) was used equipped with a narrow band pass filter (633 nm). The signal from a photodiode (FLC Electronics PIN-20) was de-convoluted with a lock-in amplifier (EG&G 7265) into $1f$ and $2f$ components to yield a retardation induced by the sample.

The dielectric permittivity was measured with the WayneKerr Precision Component Analyzer 6425, at the frequency 12 kHz, with the applied voltage amplitude ($V$) ranging from 0.1 to 5.0 V. The splay elastic constant $K_{11}$ was determined from the threshold voltage $V_{th}$ at which the director reorientation starts and thus the effective permittivity ($\varepsilon$) starts to grow as: $K_{11} = \Delta\varepsilon\, \varepsilon_0 \left(\frac{V_{th}^2}{\pi^2}\right)$. The bend elastic constant, $K_{33}$ was estimated by fitting the $\varepsilon(V)$ dependence far above the threshold voltage [30] to the formula:

$$\frac{\varepsilon(V) - \varepsilon_\perp}{\varepsilon_{II} - \varepsilon_\perp} = 1 - \frac{2}{\pi}\sqrt{1+\xi}\frac{V_{th}}{V}\int_0^1 \sqrt{\frac{1+\kappa x^2}{1+\xi x^2}}dx$$

where $\xi = \frac{\varepsilon_{II}-\varepsilon_\perp}{\varepsilon_\perp}$ and $\kappa = \frac{K_{33}-K_{11}}{K_{11}}$.

The x-ray diffraction studies that were taken with the Bruker D8 GADDS system (CuKα radiation, Goebel mirror, point beam collimator, Vantec2000 area detector) confirmed a lack of the long range order in both nematic phases. Interestingly, while for the shortest homologue (CB6O.O4) only a diffraction signal related to a half molecular length is observed, indicating that the local structure is intercalated, for the longest studied homologue (CB6O.O8) a signal matching the full molecular length is visible, suggesting a local lamellar structure (cybotactic groups) with a periodicity corresponding to the full length of the dimeric molecule.

Prior to the measurements, the quality of alignment was verified by optical microscopy. In optical measurements the transition from the nematic to the N$_{TB}$ phase was observed as a sudden 'freezing' of the texture accompanied by the quenching of 'flickering', and a few degrees below the transition the striped texture, characteristic of the periodic modulation of the optical axis in the N$_{TB}$ phase, developed. The measured optical birefringence, $\Delta n$, sharply rises at the transition from the isotropic to the nematic phase, and on further cooling follows a critical dependence of birefringence

$$\Delta n_0 = \Delta n_{max}\left(\frac{T-T_{Iso-N}}{T_{Iso-N}}\right)^{\beta_{Iso-N}} \quad (1)$$

where $T_{Iso-N}$ is the isotropic-nematic transition temperature, $\Delta n_{max}$ is the theoretical birefringence for the material with an ideal orientational order, $S = 1$, and $\beta_{Iso-N}$ is the critical exponent (Fig. 2). The critical exponent $\beta_{Iso-N}$ and $\Delta n_{max}$ for the studied materials, obtained by fitting the data to the power law, are collected in Tab. 1.

*Table 1.* Fitting parameters describing the critical temperature dependence of birefringence near the Iso-N phase transition $\Delta n = \Delta n_{max}(\frac{T_{Iso-N}-T}{T_{Iso-N}})^{\beta_{Iso-N}}$ and, deduced from birefringence, critical component of tilt fluctuations in the nematic phase on approaching the N – N$_{TB}$ phase transition: $<\theta^2_{T_c}> - <\theta^2> \sim T_r^{1-\alpha_{N-NTB}}$, $<\theta^2_{T_c}>$ is the mean square of the tilt fluctuation (in rad$^2$) at $T_c$.

|  | $\Delta n_{max}$ | $T_{Iso-N}$ | $\beta_{Iso-N}$ | $<\theta^2_{T_c}>$ | $1-\alpha_{N-N_{TB}}$ |
|---|---|---|---|---|---|
| CB6OCB | 0.31±0.01 | 152.3 | 0.15±0.02 | 0.034 | 0.95±0.03 |
| CB6O.O4 | 0.30±0.01 | 146.7 | 0.18±0.02 | 0.054 | 0.98±0.05 |
| CB6O.O5 | 0.32±0.01 | 137.1 | 0.17±0.02 | 0.049 | 1.05±0.05 |
| CB6O.O8 | 0.32±0.01 | 128.5 | 0.19±0.02 | 0.054 | 0.95±0.03 |

The order parameter was deduced from the ratio of the measured birefringence and $\Delta n_{max}$; close to the N$_{TB}$ phase $S$ reaches 0.6 - 0.7. In the CB6O.Om homologues a small jump of the orientational order, and thus birefringence is observed at the N – N$_{TB}$ phase transition, the order parameter increases by $\Delta S \sim 0.003$, indicating that the transition is very weakly first order. For the CB6OCB compound, the jump in the order parameter was below the resolution of the measurement, *i.e.* $\Delta S < 10^{-4}$, although the transition is of the first order [18]. Additionally, for the CB6O.Om homologues a small pre-transitional increase of birefringence in the nematic phase over a narrow temperature range of approximately 1 K above the N – N$_{TB}$ phase transition was detected (Fig. 2). It should be noted that for CB6OCB and the CB6O.Om homologues the birefringence departs from the critical dependence already several degrees above the N – N$_{TB}$ phase transition (Fig. 2). The decrease of the birefringence is due to the presence of strong fluctuations of the director in the nematic phase caused by the instantaneous formation of local heliconical states. Such fluctuations lead to local tilting of the molecules, and thus decrease the difference between the eigenvalues of the dielectric tensor at optical frequencies, which results in the smaller optical anisotropy of the phase [21-23]. For more quantitative analysis, the mean square fluctuation tilt angle, $<\theta^2>$, was extracted from the decrease in birefringence, assuming $\Delta n = \Delta n_0(1 - \frac{3}{2} <\theta^2>)$, where $\Delta n_0$ (see eq. (1)) is a value extrapolated from the fit of the birefringence measured over approximately 20 K temperature interval below the Iso-N phase transition, in which the heliconical fluctuations are negligible.

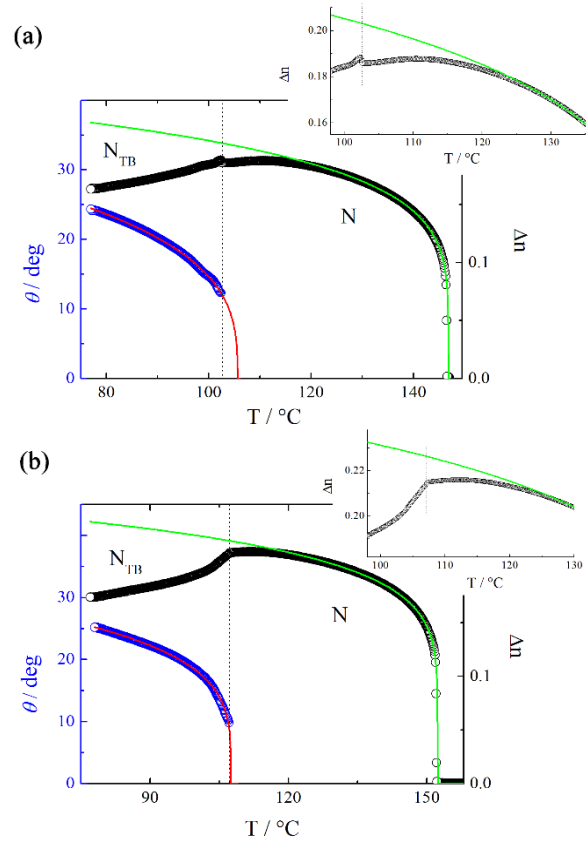

*Figure 2*. Optical birefringence (black circles) measured with red light ($\lambda = 633\ nm$) for (a) CB6O.O4 and (b) CB6OCB. The insets: a magnified part of the curves showing the deviation of $\Delta n$ from the power law dependence (green lines) near the N – N$_{TB}$ transition. Blue circles in both panels (a) and (b) show the conical tilt angle in the N$_{TB}$ phase, and the red lines represent the fit to critical dependence, $\theta = \theta_0 T_r^2$.

The fluctuations are evidently stronger for the CB6O.Om series than for CBO6CB (Fig. 3): at the transition temperature ($T_c$) the mean square fluctuation tilt angle, $<\theta^2_{T_c}>$, reaches 0.045 and 0.033 rad$^2$ for CB6O.Om and CBO6CB, respectively. The critical component of the fluctuations, $<\theta^2_{T_c}> - <\theta^2>$, is expected to follow the power law dependence $T_r^{1-a_{N-N_{TB}}}$, where $T_r$ is a reduced temperature $T_r = (T - T_c)/T_c$ and $T_c$ is the phase transition temperature, where $a_{N-N_{TB}}$ is the critical exponent for the heat capacity anomalies (Fig. 3) [24,25]. The critical exponent $1 - a_{N-N_{TB}}$ for CBO6CB and the CB6O.Om homologues was found to be close to 1 (Tab. 1). The critical fluctuations were also determined for a model $N_{TB}$ material CB7CB. In this material the range of the nematic phase is smaller and therefore the N – $N_{TB}$ phase transition is stronger first order. Although the temperature range of the critical behavior was significantly smaller and fluctuations of the heliconical tilt angle weaker, the critical exponent $1 - a_{N-N_{TB}}$ for CB7CB is close to 1, the same as for the other studied compounds (Fig. 3).

Furthermore, the birefringence data was also used to determine the conical angle in the $N_{TB}$ phase (Fig. 2), following the procedure described by Meyer [26]. It was found that 20 K below the N-$N_{TB}$ transition the conical angle reaches approximately 25 deg for both the CB6O.Om and CB6OCB compounds. The conical angle clearly decreases on approaching the nematic phase, following $\theta = \theta_0 T_r^\beta$. However, because the macroscopic tilt and tilt fluctuations cannot be separated near the phase transitions and the transition is weakly first order, the obtained critical exponents $\beta \sim 0.3$ should be treated with some caution.

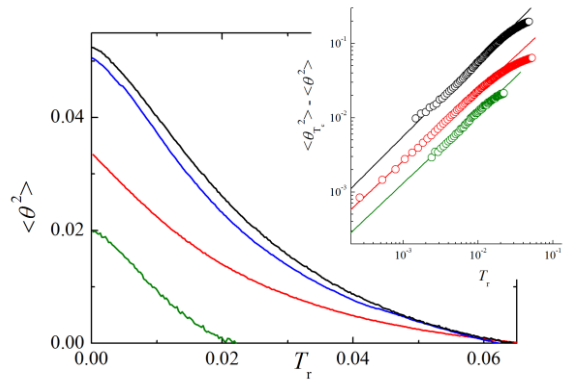

**Figure 3**. *The mean square of the fluctuations $<\theta^2>$ of the heliconical tilt angle in the N phase as a function of the reduced temperature ($T_r$) in CB6O.O4 (black), CB6O.O8 (blue) and CB6OCB (red); for comparison, the data for CB7CB are also given (green). The inset: a log-log plot of the critical component of the fluctuations, $<\theta^2_{T_c}> - <\theta^2>$, evidencing a power law dependence with the critical exponent close to 1. For clarity, the data in the inset are vertically shifted. For CB6O.O4 and CB7CB few data points very close to $T_c$ were excluded as a precritical increase of the order parameter near to the N-$N_{TB}$ transition influenced the measured values of birefringence.*

To estimate the critical exponents for the temperature dependence of the average amplitude of the fluctuation of the cone angle we use the Landau free energy introduced by Kats and Lebedev [19]. Assuming that in the nematic phase only fluctuations in the cone angle $\theta$ are present and local heliconical axis is along the $z$-direction, the Landau free energy ($F$) describing the onset of the helical order in the $N_{TB}$ phase is:

$$F = \int dV \left\{ \frac{a}{2} |\vec{\varphi}|^2 + \frac{b_3}{8q_0^2} |(-q^2 + q_0^2)\vec{\varphi}|^2 + \frac{b_1}{2} (\nabla \vec{\varphi})^2 + \frac{\lambda}{24} |\vec{\varphi}|^4 \right\} \qquad (2)$$

where $a$, $b_1$, $b_3$ and $\lambda$ are coefficients in the Landau expansion, $a$ being proportional to the reduced temperature $T_r$ and $q_0$ is the wave vector of the short pitch periodicity, at which the order parameter $\vec{\varphi}$ condenses in $N_{TB}$ phase. The magnitude of the order parameter essentially presents the heliconical tilt angle. The order parameter is a vector, its components denoting the average orientation of the long molecular axes at some point in space. The order parameter $\vec{\varphi}$ at a wave vector $q$ and the heliconical axis along the $z$-direction is

$$\vec{\varphi} = \theta_q \{\cos(qz), \sin(qz), 0\} \qquad (3)$$

where $\theta_q$ is the amplitude of this Fourier component. Plugging eq. (3) into eq. (2) the energy density is obtained:

$$\frac{F}{V} = \frac{a}{2}\theta_q^2 + \frac{b_3}{8q_0^2}(-q^2+q_0^2)^2\theta_q^2 + \frac{b_1}{2}q^2\theta_q^2 + \frac{\lambda}{24}\theta_q^4 \tag{3}$$

By keeping only the second order terms in the heliconical tilt angle and using the principle of equipartition we find

$$\langle\theta_q^2\rangle = \frac{k_B T}{\frac{a}{2} + \frac{b_3}{8q_0^2}(-q^2+q_0^2)^2 + \frac{b_1}{2}q^2}$$

Integrating over all $q$, a rather comprehensive expression for $\langle\theta^2\rangle$ is obtained, which can be Taylor expanded in terms of $a$ to give

$$\langle\theta^2\rangle = C - Da$$

where $C$ and $D$ are positive constants, depending on $b_1$ and $b_3$. As $a$ is a linear function of temperature the mean field exponent for the birefringence changes is thus 1.

Finally, the elastic constants, $K_{11}$ and $K_{33}$, were also determined for the materials studied. All the materials exhibit a positive dielectric anisotropy over the whole temperature range of the N phase, thus the elastic constant measurements were also performed in planar cells, in which an electric field was applied across the cell. For all the CB6O.Om compounds the splay elastic constant was higher than the bend elastic constant, $K_{11} > K_{33}$ (Fig. 4).

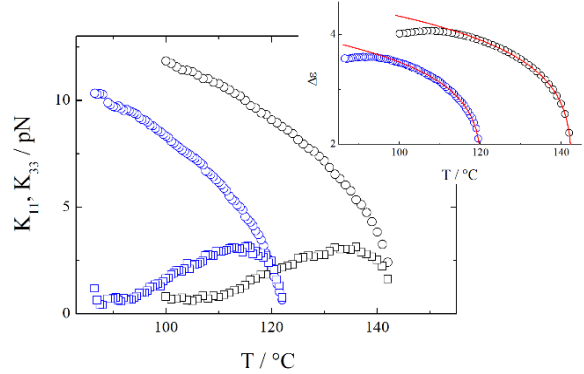

**Figure 4**. Temperature dependence of the elastic constants $K_{11}$ (circles) and $K_{33}$ (squares) in the N phase for CB6O.O4 (black), CB6O.O8 (blue). The inset: dielectric anisotropy showing deviation of $\Delta\varepsilon$ from the power law dependence $T_r^\beta$ with critical exponent $\beta = 0.17$ (red lines).

As expected, $K_{11}$ increases with decreasing temperature, following the square of the order parameter, $K_{11}\sim S^2$. By contrast, $K_{33}$ shows a non-monotonic dependence on temperature. Specifically, $K_{33}$ exhibits a clear cross-over, where in the temperature range close to the isotropic phase it increases with decreasing temperature, while on further cooling it decreases and stabilizes when reaching very small values, below 1 pN. Since $K_{33}$ is very sensitive to the molecular geometry, its anomalous temperature dependence was related to the molecular conformation distribution that changes strongly with temperature [9,27]. The plateau observed in the temperature dependence of $K_{33}$ might be the result of competing interactions: a decrease caused by conformational changes and an increase due to the helical fluctuations stiffening the elasticity of the system. On the other hand, the fluctuations clearly influenced the dielectric anisotropy of the nematic phase, presence of instantaneous helical states already several degrees above the N – N$_{TB}$ transition causes a deviation of the measured $\Delta\varepsilon$ value from the power low dependence (Fig. 4), similar to the observations in optical measurements.

To conclude, we have reported on the optical studies that showed a strong reduction of the optical birefringence on approaching the N – N$_{TB}$ phase transition, due to the instantaneous formation of helical structures in the nematic phase. A similar effect was observed in the dielectric anisotropy. The behavior was found for CBO6CB and the CBO6.Om materials, for which a well-developed fluctuation regime is induced due to the broad temperature range of the nematic phase preceding the N$_{TB}$ phase and as a result the N – N$_{TB}$ transition is very close to the second order. By comparing the measured birefringence with the extrapolation of the power-law dependence, the temperature variation of the mean square of the instantaneous cone angle, $\langle\theta^2\rangle$, and heliconical tilt angle in the N$_{TB}$ phase were determined. It was found that above the N – N$_{TB}$ phase transition temperature, the critical exponent for the mean square fluctuation tilt angle is approximately 1 and below the transition the critical exponent for tilt is approximately 0.3, both being close to values predicted by the mean field model. The difference between the measured $\beta \sim 0.3$ and theoretical mean field exponent $\beta \sim 0.5$ may be due to the low precision of the heliconical tilt angle measurements in the region close to the phase transition, where the tilt fluctuations are also strong and are thus affecting the measured birefringence. Since the critical exponent for the conical angle fluctuations correlates with the heat capacity exponent as $1 - \alpha$, our results point that $\alpha = 0$, i.e. there should be no heat capacity anomalies in the nematic phase above $T_c$, as indeed observed experimentally for both the CB7CB and CB6OCB compounds. In the calorimetric studies no heat capacity anomalies were measured in the nematic phase above the N – N$_{TB}$ transition and very weak anomalies in the N$_{TB}$ phase below the N – N$_{TB}$ transition with a critical exponent $\alpha \sim 0.5$ [9, 17, 18]. The observed mean field behavior indicates that the temperature range in which the critical scaling of fluctuations described by the xy model would only be observable over a very small temperature range, in terms of a reduced temperature at $T_r < 10^{-3}$. Such a narrow temperature range of renormalization behavior could be due to a large correlation length of helical fluctuations in the nematic phase [28], even at temperatures relatively far from the transition. The experimental access to the critical regime at $T_r < 10^{-3}$ for the systems studied is impossible due to the weakly first order character of the N – N$_{TB}$ phase transition with a coexistence range of N and N$_{TB}$ of approximately 0.1 K. Finally, it should be stressed that the amplitude of the heat capacity anomalies is directly related to the density changes, because the changes of the intermolecular distances are linked to the molecular interaction energy [29]. Apparently, the formation of the heliconical structure does not require a significant reallocation of molecules in space that would lead to the change of density. On the other hand, the tilting of the molecules influences the birefringence strongly, making the information about helical fluctuations easily accessible by optical studies.

**Acknowledgments**

This research was supported by the National Science Centre (Poland) under the grant no. 2016/22/A/ST5/00319.